# Demonstration of a low loss, highly stable and re-useable edge coupler for high heralding efficiency and low $g^2(0)$ SOI correlated photon pair sources

**JINYI DU,**[1,*] **GEORGE F.R. CHEN,**[2] **HONGWEI GAO,**[2] **JAMES A. GRIEVE,**[4] **DAWN T. H. TAN,**[2,3] **AND ALEXANDER LING**[1,5]

[1]*Centre for Quantum Technologies, 3 Science Drive 2, National University of Singapore, 117543, Singapore*
[2]*Photonics Devices and System Group, Singapore University of Technology and Design, 8 Somapah Rd, Singapore 487372, Singapore*
[3]*Institute of Microelectronics, A*STAR, 2 Fusionopolis Way, #08-02, Innovis Tower, Singapore 138634, Singapore*
[4]*Quantum Research Centre, Technology Innovation Institute, Abu Dhabi, United Arab Emirates*
[5]*Department of Physics, National University of Singapore, Blk S12, 2 Science Drive 3, 117551, Singapore*
*\*jinyidu@u.nus.edu*

**Abstract:** We report a stable, low loss method for coupling light from silicon-on-insulator (SOI) photonic chips into optical fibers. The technique is realized using an on-chip tapered waveguide and a cleaved small core optical fiber. The on-chip taper is monolithic and does not require a patterned cladding, thus simplifying the chip fabrication process. The optical fiber segment is composed of a centimeter-long small core fiber (UHNA7) which is spliced to SMF-28 fiber with less than -0.1 dB loss. We observe an overall coupling loss of -0.64 dB with this design. The chip edge and fiber tip can be butt coupled without damaging the on-chip taper or fiber. Friction between the surfaces maintains alignment leading to an observation of ±0.1 dB coupling fluctuation during a ten-day continuous measurement without use of any adhesive. This technique minimizes the potential for generating Raman noise in the fiber, and has good stability compared to coupling strategies based on longer UHNA fibers or fragile lensed fibers. We also applied the edge coupler on a correlated photon pair source and observed a raw coincidence count rate of 1.21 million cps and raw heralding efficiency of 21.3%. We achieved an auto correlation function $g_H^{(2)}(0)$ as low as 0.0004 at the low pump power regime.

## 1. Introduction

Silicon photonics has achieved remarkable progress in the past decade. It offers a potentially cost-effective and scalable platform for developing silicon-based optical circuits. Amongst the various platforms available for integrating optical and electrical devices, silicon on insulator (SOI) is a good candidate. Besides a mature fabrication process, the high refractive index contrast between silicon and silicon oxide leads to small optical mode volumes that enable highly compact photonic circuits with small footprints [1,2].

The tightly confined optical modes, however, pose a challenge for coupling into optical fibers. This is an issue in applications which are sensitive to loss, for example, in quantum technologies where quantum correlated light generated on chip must be routed to off-chip instruments. The high coupling losses arise from the mode mismatch between the fibers and SOI waveguides [3-5], For example, the mode field diameter of a standard single mode fiber SMF-28 is 10.4 μm at 1550 nm but less than 1 μm in a silicon waveguide.

To solve challenges associated with mode mismatch, the fiber mode could be reduced, the waveguide mode could be enlarged, or both could be done at the same time. Many methods have been employed to convert the fiber mode with the help of lensed fibers [2,6-8] and HNA fibers [3,4,9,10]. To enlarge the waveguide mode, an inverse taper often requires an additional structure to assist the mode expansion. For example, mode expansion has been demonstrated using a two-layer silicon nitride (SiN) structure [3,4], a multi-rod solution [9], multi-fork tapers [5,6,7,11], sub-wavelength grating SiN layers [4,6], and 3-dimensional tapers [12,13]. These solutions can realize high coupling efficiency but are complex to fabricate and can be difficult to integrate into a standard foundry workflow.

In this paper, we present a fiber coupling strategy that can approach the minimum predicted loss when collecting photons from a silicon photonics waveguide. In this approach, a centimeter-long UHNA7 fiber is spliced to an SMF-28 fiber with less than -0.1 dB splicing loss, converting the fiber mode from 10.4 μm to 3.2 μm. On the chip side, a two-stage inverse taper is fabricated to expand the chip mode to reduce the mode mismatch with that of the UHNA7 fiber.

Besides coupling efficiency, long-term stability is also an essential metric for chip couplers. To realize high stability, refractive index matching oil [3,14] and epoxy [14] are widely used to increase the alignment tolerance or permanently fix the fiber and chip. However, these methods introduce contamination and may be irreversible. To avoid these issues, efforts have been made to design the coupler to tolerate a mismatch in the alignment between chip and fiber [3,11]. State-of-the-art designs can provide a -1 dB loss for a displacement of 3 μm on the x and y axes [3,5,9,12-16]. This still necessitates using an active feedback alignment stage [2] for long-term measurements. In this work, we show that pushing the cleaved fiber tip against the chip facet allows friction between them to maintain a high coupling efficiency for extended periods. This method does not harm the coupler or fiber, and importantly, it avoids contaminating the chip with non-removable optical glue, thus allowing for the chip's reuse.

Quantum correlated sources are essential components in quantum communication, and they can be engineered by pumping a short silicon waveguide which has strong $\chi^3$ nonlinearity. In the literature, silicon photonics quantum correlated sources have a lower performance compared to sources using nonlinear optical fiber or crystals such as PPLN and PPKTP in terms of observed coincidence count rate, measured raw heralding efficiency [17-20]. The main reason is the high loss fiber-chip optical coupling in most silicon integrated sources [21,22].

In this paper, we applied the novel edge coupler on a correlated photon pair source and enhanced the measured brightness and heralding efficiency. The high stability of the edge coupler enables data integration over extended hours, ensuring that the auto correlation function $g_H^{(2)}(0)$ measurement remains stable even in a low pump power regime. Our work shows that the performance of integrated photonics quantum correlated sources can be close to the bulk crystal sources based on SPDC and offer considerable advantages in terms of size, ease of phase matching, and the potential for scalable mass production.

## 2. The edge coupler designs

In our design, the definition of coupling efficiency is composed of several factors:

$$\eta_{coupling} = \eta_{overlap}\eta_{conversion}\eta_{splicing}\eta_{reflection}$$

where $\eta_{overlap}$ is the mode overlap between the on-chip coupler and fiber, $\eta_{conversion}$ is the efficiency of adiabatic mode conversion for the on-chip waveguide, $\eta_{splicing}$ is the SMF-28 and UHNA7 fiber splicing efficiency, while $\eta_{reflection}$ is the reflection of the air gap at the chip facet due to the refractive index contrast.

The coupling strategy uses two matching mode converters. One mode converter is on the photonic chip, and the other is on the tip of optical fiber. The mode overlap $\eta_{overlap}$ is defined as:

$$\eta_{overlap} = \frac{\left|\int E_1 E_2 dA\right|^2}{\int |E_1|^2 dA \int |E_2|^2 dA}$$

where $E_1$ is the complex electric field amplitudes of the optical fiber mode, while $E_2$ is that of the silicon waveguide facet mode [23].

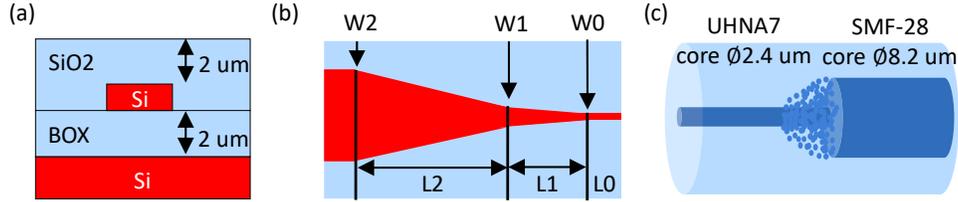

Fig. 1. (a). The cross-sectional view of the silicon chip. The silicon waveguide is on the top of 2 μm buried oxide (BOX) layer and covered by 2 μm SiO2 upper cladding. (b). Top view of on-chip mode converter with BOX layer. Not shown is the upper cladding. (c). The fiber mode converter. The tapered region adiabatically converts the UHNA7 fiber mode to that of SMF-28.

## 2.1 The on-chip mode converter

The structure of the on-chip coupler is demonstrated in Fig.1. (a) and (b). The on-chip mode converter has a 15 μm input waveguide (L0) that serves as a buffer to protect the inverse taper during the cleaving process. The length of the first stage taper is 50 μm (L1). This first stage expands the waveguide width from W0 (0.16 μm) to W1 (0.27 μm). The second stage taper has a length of 100 μm (L2) and expands the waveguide width from W1 to W2 (0.65 μm). The width W2 is that of the straight waveguide. The three main parameters, the first stage taper tip width (W0) and length (L1), and the second stage taper's length (L2) are optimized by the EME solver of ANSYS Lumerical. The first stage taper is more gradual than the second stage taper because of the tradeoff between an efficient adiabatic mode conversion and the need to maintain a small physical footprint. The silicon waveguide height is 0.25 μm and covered with 2 μm thick SiO$_2$ upper cladding.

## 2.2 The fiber mode converter

An important consideration in the coupling strategy is to obtain the smallest possible mode field diameter at the tip of the collecting fiber. The UHNA7 fiber offers the minimal mode size from a list of available off-the-shelf fibers and can be used on the fiber tip for initial collection. The UHNA7 fiber is then spliced to a regular SMF-28 fiber for onward routing. The structure of the fiber coupler is demonstrated in Fig. 1. (c). To optimize the shape of the tapered region and reduce losses, a fiber splicer is used to perform connection between the two fibers.

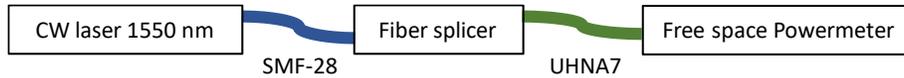

Fig. 2. The in-situ setup for checking loss during the splicing process. The fiber splicer is Vytran 3000 with a graphite filament.

A low loss splice between the UHNA7 and SMF-28 fibers is typically achieved by using low filament power and long splice time to diffuse the dopants within the UHNA7 fiber. After diffusion, the fiber modes can be adiabatically converted [24]. If the splice time is too lengthy, it will melt the splicing region and result in higher loss. The splicing between UHNA7 and SMF-28 was discussed in [24], but such presented parameters are typically machine specific (Vytran FFS2000WS).

To optimize the parameters for different machines, it is preferable to perform an in-situ monitoring of the loss during the splicing process. The concept is shown in Fig. 2. A low power 1550 nm laser beam is coupled into the SMF-28 section, and the transmitted power is measured in real time at the output of the UHNA7 fiber. The splicing is performed with a Vytran 3000 glass processor, and Fig. 3 shows the splicing losses versus splice duration for different filament powers.

This method of in-situ loss monitoring allows for immediate filament power adjustment based on the observed changes in the slope of the splicing loss curve. As a result, the optimal splicing parameter settings can be obtained with fewer attempts. The data presented in Fig. 3 indicates that splicing losses lower than -0.1 dB are achievable with a splice time of 60 seconds and a filament power of 58 W.

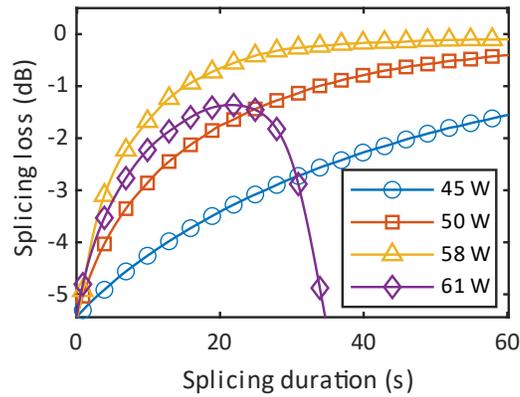

Fig. 3. The splicing loss during the splicing process for different filament powers

The heavily doped UHNA7 fiber results in a high refractive index contrast, but also enables more Raman light generation in the presence of an intense pump field [25,26]. This is a problem for noise sensitive applications such as quantum experiments that generate correlated photons via spontaneous four wave mixing (SFWM) as the Raman noise reduces the signal-to-noise ratio (SNR). To minimize the presence of Raman noise, the UHNA7 fiber in this work is cleaved to a distance of approximately 1 cm after splicing. This residual distance is dependent on the cleaver available.

## 3. Experimental results

### 3.1 Coupling loss testing

Fabrication of the devices is performed on silicon-on-insulator wafers with a silicon thickness of 250 nm and buried oxide thickness of 3 μm on a silicon substrate. Electron-beam lithography is used to pattern the devices followed by reactive ion etching. A 2 μm layer of silicon dioxide is deposited after etching using plasma enhanced chemical vapor deposition.

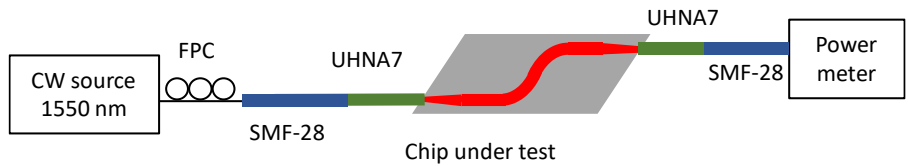

Fig. 4. The experimental setup for characterization of the chip insertion loss. The fiber polarization controllers (FPC) at the input arm optimize the polarization of the input beam to reach the highest coupling efficiency. The UHNA7 fibers are labeled green. Two on-chip mode converters are connected with short waveguides and are displaced by 200 μm to minimize coupling via scattered light between the input and output fibers.

In the chip layout, there are waveguides with different lengths ranging from 3 mm to 22 mm. The purpose of the various waveguide lengths is to linearize the measurement result and subtract the linear propagation loss.

As shown in Fig. 4, a 1550 nm continuous wave (CW) low power laser is launched into the chip from an optical fiber using a UHNA7 tip that matches the mode of the on-chip coupler. The chip is stably mounted on a copper cube, and the fiber-chip alignment is done with a six-axis piezo stage (Thorlabs MAX609). Light from the chip is collected via another pair of mode-matched on-chip couplers and UHNA7 tip, and the output power is recorded.

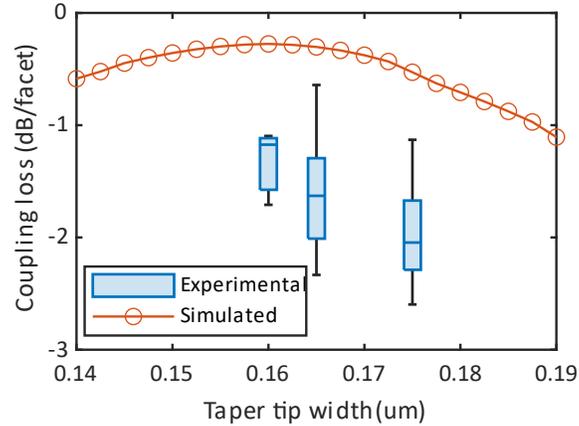

Fig. 5. Comparison of the coupling losses predicted by the model (solid line) and the observed experimental data (boxplots). The experimental data comes from 23 individual couplers on 3 wafers and each wafer has a different taper tip width. The black bars show the maximum and minimum observed coupling loss, while the blue boxes demonstrate the range within the upper quartile and lower quartile.

A summary of the observed coupling losses is provided in Fig. 5. The minimum measured coupling loss is -0.64 dB at 0.165 μm tip width. For the coupler group with 0.16 μm tip, the median of the data set is -1.17 dB. Improvements in the fabrication process to mitigate process related bias could help to reduce the standard deviation of the observed coupling losses across different devices. One possible area for investigation is the length of the buffer region, L0, as the narrow waveguides bring more transmission loss compared with wide waveguides [27]. Another potential area for study in the future is to further reduce the variation in the coupler performance by exploring methods for mitigating fabrication related bias. Silicon photonics foundries which provide wafer scale fabrication of devices may be one possible option to fabricate large numbers of devices with low die to die variations as they have greater control in the CMOS process.

### 3.2 Long-term stability

Long-term stability is another parameter to be investigated when evaluating the performance of edge couplers. In this work, the chip is firmly glued on a metal mount and the two cleaved fiber tips are pressed against the chip as shown in Fig. 6. This arrangement ensures that the chip remains stationary when subjected to tension from the fibers.

The friction between fiber and chip can help the coupler maintain coupling against the mechanical vibrations. When scanning the fiber holder position with a closed loop piezo stage, the fiber tip bent and utilized this deformation to compensate for fiber holder movements. The edge coupler kept ±0.1 dB extra loss within ±10 μm fiber holder movement in the x and y axes. This property ensures high coupling stability over a long period of time, which is useful in quantum experiments.

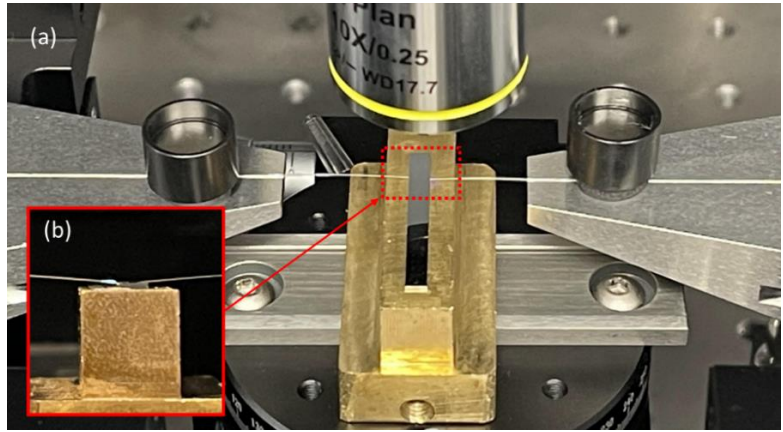

Fig. 6. Demonstration of friction-assisted edge coupler. (a). The chip is glued to a metal mount. Two fibers, each mounted on a fiber holder, are positioned to push against the chip from opposite sides. (b). Side view of the chip: the fibers are intentionally slightly bent, allowing their deformation to compensate for any misalignment.

The experimental setup to characterize long-term stability is shown in Fig. 7. (a). Power meter 1 monitors the input light power by inserting a 50:50 beam splitter into the optical path. The fiber chip coupling is conducted with the friction-assisted edge coupler. Power meter 2 is connected to the output fiber to measure the output power from the chip. The insertion loss attributed to a pair of edge couplers is determined by comparing the measured input power to the output power. The raw data is recorded by averaging the power every three minutes and shown in Fig. 7. (b) and (c) with scatter plots. In Fig. 7. (c), the output power's raw data fluctuated in the ±0.1 dB range over the ten-day measurement period, and the 1-day moving average value fluctuated from -0.02 dB to 0.01 dB, indicating excellent long-term stability without using any adhesive.

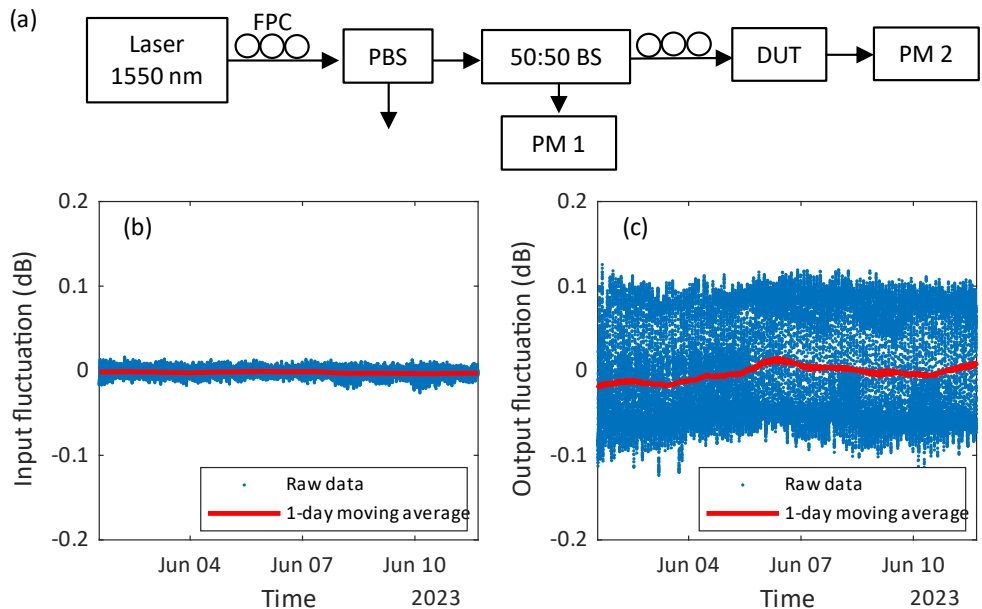

Fig. 7. Experimental observation of long-term coupling stability. (a). The experimental setup. FPC: fiber polarization controller. BS: beam splitter. DUT: device under test. PM: power meter. (b). The chip input power fluctuation in ten days measured by Power meter 1. (c). The chip output power fluctuation in ten days measured by Power meter 2.

*3.3 State-of-the-art comparison*

**Table 1. Comparison of various edge couplers**

| Structure | Fiber type | TE coupling loss (dB) | Long-term stability | Ref. |
|---|---|---|---|---|
| Suspended taper | SMF with oil and epoxy | -1.4 | NaN | [14] |
| SiN layer-assisted taper | HNA (6.5 μm MFD) with oil | -0.35 | NaN | [3] |
| Cantilevered converter | SMF with oil | -1.5 | NaN | [13] |
| Multi-rod structure | HNA (6.6 μm MFD) | -0.5 | NaN | [9] |
| Arrayed waveguides | SMF | -1.7 | NaN | [5] |
| 3D polymer taper | SMF | -1 | NaN | [12] |
| SION lens structures | SMF | -4 | NaN | [15] |
| Fork shape edge coupler | 6 μm MFD lensed fiber | -1.25 | NaN | [7] |
| Meta-trident coupler | 2.5 μm MFD lensed fiber | -0.5 | NaN | [6] |
| SiN SWG assisted taper | SMF | -0.42 | NaN | [4] |
| Silicon taper | short section UHNA7 | -1.5 | NaN | [10,24] |
| Multi-layer LNOI taper | UHNA fiber | -0.54 | ±0.01 dB for 2.5 hours with UV glue | [28] |
| Two-stage silicon taper | centimeter long UHNA7 | -0.64 | ±0.1 dB for ten days without glue | This work |

Table. 1 compares our work with various edge couplers reported in the literature. The coupler reported in this work is amongst the highest coupling efficiency couplers and reaches remarkable long-term stability by using only butt coupling.

## 4. Correlated photon pair source with novel edge coupler

Correlated photon pairs can be generated via spontaneous four wave mixing (SFWM). In the SFWM process, two pump photons at frequency $\omega_{pump}$ are annihilated, while two new photons are generated at frequencies $\omega_{signal}$ and $\omega_{idler}$ respectively, according to the expression for energy conservation, $2\omega_{pump}=\omega_{signal}+\omega_{idler}$.

*4.1 Enhancing brightness and heralding efficiency*

To show the impact of our edge coupler on integrated correlated sources, we pumped an 8 mm silicon waveguide sample with 1550.12 nm continuous wave (CW) laser and performed coincidence measurements.

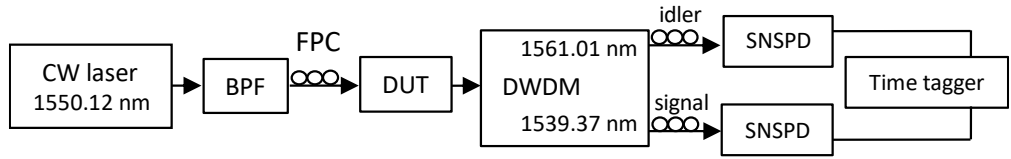

Fig. 8. The experimental setup for characterizing photon pair generation. BPF: band pass filter. FPC: fiber polarization controller. DUT: device under test. DWDM: Dense wavelength-division multiplexer. SNSPD: Superconducting nanowire single-photon detector.

As shown in Fig. 8, the pump laser propagates through a bandpass filter with 150 dB extinction ratio to remove undesired noise components from the pump spectrum. Subsequently, the filtered pump is coupled into the silicon waveguide via the novel edge coupler described above. The photon collection is also done with the same method. The waveguide employed between the two couplers is a ridge waveguide with a 2 μm upper SiO2 cladding. The geometrical dimensions are 0.25 μm in height and 0.65 μm in width.

The DWDM filters employed for pump rejection have 0.8 nm flat-top transmission band at their center wavelengths, set at 1561.01 nm and 1539.37 nm, respectively. These filters have an extinction ratio greater than 100 dB for pump photons, but this high level of rejection is maintained only when the wavelengths are positioned at least 5 nm away from the center wavelength. As the wavelength deviates closer than this 5 nm threshold from the center, the extinction ratio decreases, indicating a reduction in the filter's effectiveness at blocking pump photons. This characteristic underlines the trade-off in signal and idler photon collection wavelength selection, balancing the photon pair generation rate against the heralding efficiency. A narrower wavelength separation would increase the photon pair generation rate but significantly reduce the heralding efficiency due to the presence of unfiltered pump photons. Conversely, wider separation would preserve high heralding efficiency but reduce the photon pair generation rate due to the increased challenge of meeting phase matching conditions.

The system loss in each channel (DWDM losses and fiber connector losses) is -0.6 dB for the idler channel and -0.8 dB for the signal. Photons are finally detected using Superconducting nanowire single-photon detectors (SNSPDs) with approximately 38% detection efficiency for idler channel and 61% for signal. The methodology for calibrating the detection efficiency is detailed in [29]. Since the SNSPD is polarization sensitive, polarization controllers are applied at both channels to optimize the detection efficiency. The time tagger calculates the coincidence counts of the two channels with different coincidence windows.

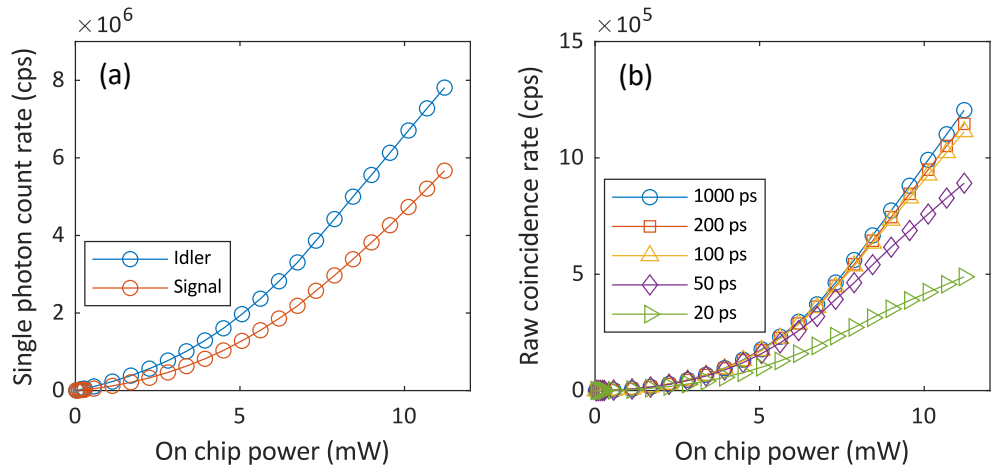

Fig. 9. (a). Single photon count rate of each channel measured by SNSPDs with coincidence window 1 ns. (b). The coincidence count rates calculated by time tagger with different coincidence windows.

Figure 9. (a) shows the measured single photon count rates of signal and idler channel versus on chip power. The signal and idler values differ slightly because the filtering loss is unbalanced in signal and idler channels. In Fig. 9. (b), the coincidence count rate increases quadratically because two pump photons are needed in the SFWM process. The maximum measured coincidence count rate is 1.21 million cps at 11.2 mW on chip power. Heralding efficiency is defined as $\eta_{s(i)} = \frac{N_{coin}}{N_{i(s)}}$ where $N_{coin}$ is the measured raw coincidence rate of signal and idler channels and $N_{i(s)}$ is the rate of the idler (signal) photon.

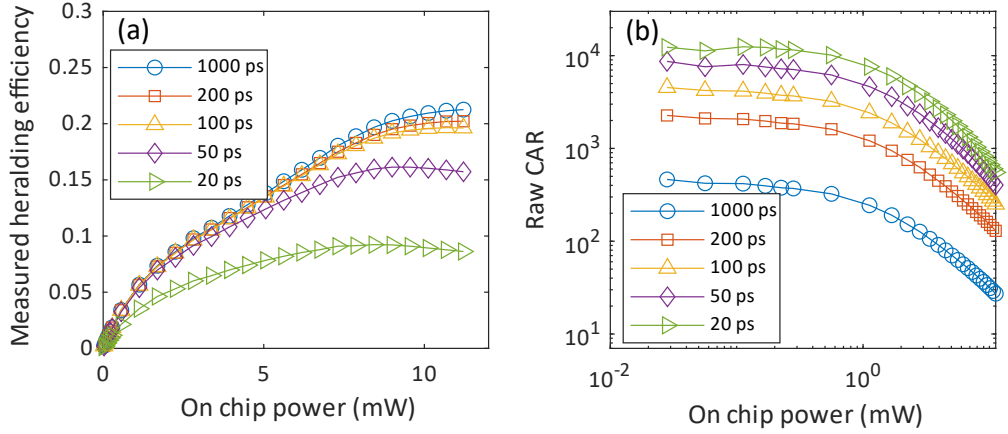

Fig. 10. (a). The raw measured heralding efficiency of the idler channel $\eta_i = \frac{N_{coin}}{N_s}$ with coincidence count at different coincidence windows. (b). Raw coincidence to accidental ratio measured with different coincidence windows.

As shown in Fig. 10. (a), the off-chip measured heralding efficiency reaches 21.3% at the maximum on chip power. The bottleneck here is the linear transmission loss and nonlinear absorption in the silicon waveguide and the noise photons generated in the optical fibers and SiO2 cladding. In Fig. 10. (b), the raw coincidence to accidental ratio (CAR), $CAR = \frac{C_{raw}}{A_{raw}}$, reaches 12,459 at 0.11 mW with 20 ps coincidence window. The coincidence data are averaged over 100 second for on chip power less than 1.7 mW, and 10 second at higher powers.

## 4.2 Improving the auto correlation function $g_H^{(2)}(0)$

In the setup of Fig. 8, when using signal photon as the heralding arm and inserting a beam splitter to the idler arm, the antibunching characteristic can be evaluated with auto correlation function $g_H^{(2)}(\tau)$. The single photon sources' $g_H^{(2)}(\tau)$ exhibit a clear antibunching dip at zero delays. This is because of the intrinsic single photon emission property and $g_H^{(2)}(0)$ can be used to quantify the multi-photon event of a single photon source. The auto correlation function of photon pair sources is defined as $g_H^{(2)}(\tau) = \frac{N_{s,i1,i2(\tau)} N_s}{N_{s,i1} N_{s,i2(\tau)}}$, where $N_{s,i1,i2(\tau)}$ is the coincidence count rate of the signal channel and the two idler channels. $N_s$ is the signal channel count rate while $N_{s,i1}$ ($N_{s,i2(\tau)}$) is the coincidence count rate of the signal channel and idler1 (idler2 delayed with $\tau$). To compare the multi photon emission across varying pump powers and coincidence windows, the delay $\tau$ should be set to 0, where the antibunching is strongest [30-32]. Figure 11 shows the raw $g_H^{(2)}(0)$ varies across on chip power and coincidence window. With an increase in on chip power, there is a corresponding rise in $g_H^{(2)}(0)$ values, indicative of a higher chance of multiphoton events. This phenomenon can be attributed to the enhanced efficiency of the spontaneous four wave mixing (SFWM) process at higher pump intensities, leading to an increased rate of multiphoton pair generation. The typical data collection times for the photon coincidences are 20 seconds to 20 minutes. However, at low on chip power (< 1.7 mW), narrow coincidence window (< 60 ps), the integration times are 1 hour. The measurements are repeated three times at each on chip power-coincidence window combination to minimize the measurement uncertainty.

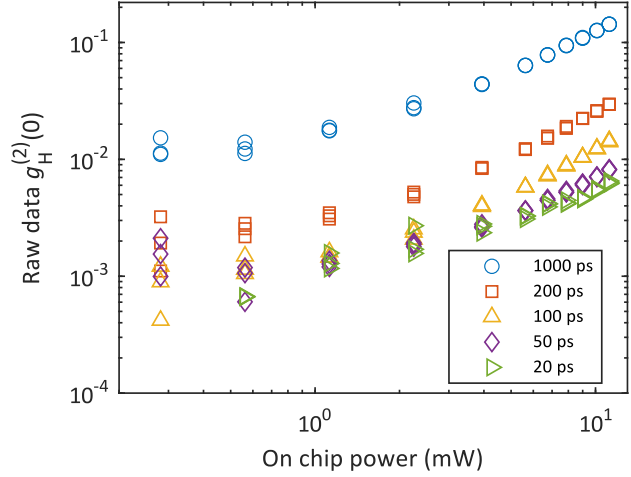

Fig. 11. Measured raw heralded self-correlation function values with different on chip power and coincidence window. Three separate data sets are collected at each on chip power-coincidence window setting combination. (Some measurement data have very low fluctuation so the data points overlap and may not be distinguishable.)

The minimum measured $g_H^{(2)}(0)$ was 0.0004 with 100 ps coincidence window at lowest on chip power 0.28 mW. The $g_H^{(2)}(0)$ for 200 ps at the highest pump is 0.03. Under this condition, the heralding efficiency is 20.2%, while the coincidence count rate is 1.15 million cps.

There is a tradeoff between the four parameters: coincidence rate, heralding efficiency, raw CAR, and raw $g_H^{(2)}(0)$. This is because the Raman noise generated from optical fibers, and SiO2 chip cladding is linearly related to the pump power while the photon pairs generated from SFWM process quadratically increase with the pump power. So, the heralding efficiency at low pump power regime is not able to be as high as at high pump power regime. However, the chance of multi photon emission is also lower at the lower pump regime and the CAR and $g_H^{(2)}(0)$ performances are better. Table 2 shows our correlated photon pair source's performance with different settings. The uncertainties in the $g_H^{(2)}(0)$ values are estimated from the deviations observed across three measurements data points, as shown in Fig. 11.

Table 2. Comparison of source performances across different settings

| On chip power (mW) | Coincidence window (ps) | Raw coincidence rate (cps) | Raw HE (including detector 38%) | Raw CAR | Raw $g_H^{(2)}(0)$ |
|---|---|---|---|---|---|
| 11.2 | 1,000 | 1.21 M | 21.3% | 27 | 0.1432(1) |
| 11.2 | 200 | 1.15 M | 20.2% | 129 | 0.0296(1) |
| 11.2 | 100 | 1.11 M | 19.6% | 251 | 0.0140(1) |
| 6.7 | 50 | 318 k | 14.6% | 882 | 0.0049(1) |
| 0.56 | 50 | 1,600 | 3.3% | 6,191 | 0.0006(2) |
| 0.28 | 100 | 417 | 1.9% | 3,677 | 0.0004(4) |
| 0.11 | 20 | 41 | 0.5% | 12,459 | NaN |

Table. 3 compares our work with various silicon correlated photon pair sources reported in the literature. Employing a low loss waveguide and reducing on chip linear loss would optimize the heralding efficiency further [1]. Apart from low loss waveguides, a high speed SNSPD capable of detecting high rates of single photons without sacrificing detection efficiency is helpful for increasing the measured raw heralding efficiency [29].

Table 3. Comparison of various silicon chip correlated source

| Type | Coupler (coupling loss) | Detector (efficiency) | Measured raw heralding efficiency | Ref. |
|---|---|---|---|---|
| Ring resonator | Edge coupler (-3.5 dB) | SNSPD (65%) | ~2.6% | [2] |
| Ring resonator | Edge coupler (-1.5 dB) | Superconducting detector (7%) | 0.6% | [10] |
| Spiral waveguide | Edge coupler (-3 dB) | SNSPD (85%) | ~7.2% (300 kHz off chip pair rate and 58 MHz on chip pair rate) | [21] |
| Ultra-low loss Multimode waveguide | Grating coupler (-6.6 dB) | SNSPD (76.4%) | 12.6% | [1] |
| Dual-interferometer-coupled microring | Grating coupler (-3.6 dB) | SNSPD (73-75%) | ~8.0% | [22] |
| Single mode waveguide | Edge coupler (-0.64 dB) | SNSPD (38%) | 21.3% | This work |

## 5. Conclusion

In conclusion, an effective coupling strategy for the SOI platform has been demonstrated. A high coupling efficiency and stability are realized without using a complex patterned overlayer for the on-chip coupler. Furthermore, the technique did not require refractive index matching oil or an active alignment mechanism. The best performing chip-fiber coupler demonstrated a loss of only -0.64 dB and provided ±0.1 dB alignment stability over ten days without any glue or active alignment stage. The same strategy could be investigated for other photonic chip platforms, for example in SiN or for waveguides that are being investigated in the rapidly developing thin-film LiNbO3 platform. Using UHNA7 fiber as the collecting fiber tip has been shown to be effective. These small core fibers can be spliced efficiently to regular SMF-28 fibers by performing in-situ observations of the transmission power during the actual splicing process.

With regards to the consistency of the coupler performance, it is noted that the yield of working couplers needs to be increased. From the experiment, only one of the couplers out of the batch of 23 devices achieved this high coupling efficiency. Although the "hero" device provided evidence of the success of the technique, it would be preferable if more of the fabricated devices could achieve a similar level of performance. Currently, the hypothesis is that ultra-low coupling losses similar in value to -0.64 dB coupling losses achieved in this work may be more consistently achieved across the fabricated devices by minimizing fabrication related variations or bias. In the future, fabrication of the designed couplers in a CMOS photonics foundry could provide economies of scale, improved repeatability and provide a higher yield of low loss couplers. One consideration is the resolution available in such foundries, since the taper width required is 165 nm whereas many foundries for silicon photonics adopt 193 nm lithography.

In terms of the applicability of the high efficiency couplers to quantum photonics, we have demonstrated a chip based high heralding efficiency source whose performance is close to that based on SPDC with bulk crystals [16-19]. The improvement was realized by only using a low loss edge coupler. The measured heralding efficiency reported in this work is amongst the highest in Table. 2. but has a very simple design and is passively stable. Silicon chip photonics offer inherent advantages in terms of scalability and reproducibility. Unlike bulk crystal systems that may require precise alignment to fulfill phase matching condition, silicon photonics can be precisely fabricated using mature semiconductor manufacturing processes. This not only ensures consistent performance across chips but also makes them more reliable for high-volume mass production.

## 6. Back matter

### 6.1 Funding

This research is supported by the National Research Foundation, Singapore, and A*STAR under its CQT Bridging Grant and Quantum Engineering Programme (Award No. NRF2021-QEP2-01-P02). D. Tan acknowledges funding from the Ministry of Education Tier 2 Grant (Award No. MOE2019-T2-2-178) and the National Research Foundation, Singapore, and A*STAR under its QEP 2.0 Programme (Award No. NRF2022-QEP2-01-P08).

### 6.2 Acknowledgments

The authors would like to thank Prof. Shayan Mookherjea, Dr. Xiaoxi Wang, Dr. Chaoxuan Ma for their contribution at the early stage of the project. We thank Prof. Xiaoying Li for the fruitful discussion. We thank Dr. Chetan Sriram Madasu and Dr. Anindya Banerji for their comments on the manuscript.

### 6.3 Disclosures

The authors declare no conflicts of interest.

### 6.4 Data availability

Data underlying the results may be obtained from the authors upon reasonable request.